\documentclass[conference]{IEEEtran}

\usepackage{times,amsmath,epsfig}
\PassOptionsToPackage{hyphens}{url}
\usepackage{url}
\usepackage{hyperref}
\usepackage{multirow}
\usepackage{caption}
\usepackage{footnote}
\usepackage{graphicx}
\usepackage{amssymb}
\usepackage{array}
\usepackage{color}
\usepackage{verbatim}
\usepackage{lmodern}

\usepackage[ruled]{algorithm2e}

\IEEEoverridecommandlockouts

\newcolumntype{C}[1]{>{\centering\arraybackslash}m{#1}}

\begin{document}

\title{Mining Anonymity: Identifying Sensitive Accounts on Twitter}

\author{
\IEEEauthorblockN{Sai Teja Peddinti\IEEEauthorrefmark{1}, Keith W. Ross\IEEEauthorrefmark{1}\IEEEauthorrefmark{2}, Justin Cappos\IEEEauthorrefmark{1}}
\IEEEauthorblockA{\IEEEauthorrefmark{1} Dept. of Computer Science and Engineering, NYU}
\IEEEauthorblockA{\IEEEauthorrefmark{2} NYU Shanghai}
\IEEEauthorblockA{psaiteja, keithwross, jcappos@nyu.edu}
\thanks{A shorter 4-page version of this work has been published as a poster in the International AAAI Conference on Web and Social Media (ICWSM), 2016~\cite{PeddintiICWSM}. This work was supported in part by the NSF (under grant CNS-1318659).}
}

\maketitle

\begin{abstract} 

We explore the feasibility of {\em automatically} finding accounts that publish sensitive content on Twitter. One natural approach to this problem is to first create a list of sensitive keywords, and then identify Twitter accounts that use these words in their tweets. But such an approach may overlook sensitive accounts that are not covered by the subjective choice of keywords. In this paper, we instead explore finding sensitive accounts by examining the percentage of anonymous and identifiable followers the accounts have. This approach is motivated by an earlier study showing that sensitive accounts typically have a large percentage of anonymous followers and a small percentage of identifiable followers. 

To this end, we first considered the problem of automatically determining if a Twitter account is anonymous or identifiable. We find that simple techniques, such as checking for name-list membership, perform poorly.  
We designed a machine learning classifier that classifies accounts as anonymous or identifiable. 
We then classified an account as sensitive based on  the percentages of anonymous and identifiable followers the account has. We applied our approach to approximately 100,000 accounts with 404 million active followers. The approach uncovered accounts that were sensitive for a diverse number of reasons. These accounts span across varied themes, including those that are not commonly proposed as sensitive or those that relate to socially stigmatized topics. To validate our approach, we applied Latent Dirichlet Allocation (LDA) topic analysis to the tweets in the detected sensitive and non-sensitive accounts. LDA showed that the sensitive and non-sensitive accounts obtained from the methodology are tweeting about distinctly different topics. We further confirmed that independent human evaluators generally agree with our automated sensitive account classification results. Our results show that it is indeed possible to objectively identify sensitive accounts at the scale of Twitter.

\end{abstract}

\IEEEpeerreviewmaketitle

\section{Introduction}
\label{sec:introduction}

Developing policy for online privacy is complex because users' preferences are highly contextual and vary based on the type of content being shared and the sender's relationship to the recipient~\cite{nissenbaum2009privacy}. Introducing appropriate privacy controls necessitates defining and identifying sensitive content, that is, content that needs special consideration and protection. 
However, there is a clear disparity among the legal and data protection authorities and the online service providers about what constitutes sensitive content~\cite{aboutCNIL,microsoftadpolicy,Facebookadguidelines,Googlepersonalinfo}.
This lack of a universal definition makes it hard to simply enumerate a list of sensitive content categories. 
In addition, due to the contextual differences, sensitive content categories identified for one application might not directly translate to another. These challenges complicate the process of identifying online content users would deem sensitive.

In this paper we consider identifying sensitive content on Twitter, to help design novel privacy policies and controls for online social media applications. Specifically, we seek to 
develop an efficient automated means for identifying accounts that tweet sensitive content.  An automated means for identifying sensitive accounts is of interest
for several reasons:
\begin{itemize}
\item Social networks suffer from intentional service abuses and illegal activities, such as spreading child pornography, weapons trafficking, or sales of narcotics.~\cite{twitterchildporn,facebookguns}.  Although Twitter by itself does not regulate content~\cite{twittermediapolicy}, legal authorities could benefit from an automated sensitive account detection system, helping them sift through vast amounts of Twitter data and narrowing their investigation targets.
\item  Twitter does not regulate content in tweets, 
but it does require that content that might be considered sensitive -- such as nudity, violence, medical procedures -- be appropriately tagged as such~\cite{twittermediapolicy}. Currently, Twitter primarily relies on highly irregular (but effective) crowd reporting to detect any policy violations~\cite{twitterpornographypolicy}, but its reach is limited due to human involvement. Automated sensitive account identification can provide an additional signal to detect policy violations and help maintain the health of the Twitter ecosystem.
\item As we show in this work, many Twitter accounts addressing self-help in stigmatized areas such as obesity, anorexia, and depression will be identified by our methodology. This reaffirms that content sensitivity is quite nuanced and is not restricted to the usual suspects.
Once these sensitive self-help sites are identified, they can be more easily shared with those seeking or providing help.
\item Finally, the issue of content sensitivity has become a fundamental question for modern social media~\cite{PeddintiCOSN2014,correa2015whisper,Peddinti2014,WangNKALC2011}. Novel means for identifying sensitive content can shed significant insight on contemporary social media, and aid in updating privacy features and policies. 
\end{itemize}

\subsection{Traditional Approaches}

One natural approach of identifying sensitive accounts is to specify categories of sensitive topics, and then identify words that commonly occur when discussing these topics. We could then search the tweets to see which accounts employ
these ``sensitive words'' in both tweet words and hash tags. This approach has the drawback of being highly \textit{subjective}, in that it relies on humans to define sensitive topics and words. Also, it is shown that humans can easily miss many topic categories due to subjectivity~\cite{Peddinti2014}.
In machine language terminology, this sensitive-word based approach tends to ``overfit'' the selected keywords and does not generalize well across all possible topics.

A second approach would be to apply automatic topic identification techniques, such as Latent Dirichlet Allocation (LDA), on tweets to identify sensitive topic themes. We can then identify accounts that relate to these sensitive themes. 
However, automatic topic identification techniques are highly resource intensive. Processing tweets of a few million users requires powerful machine clusters with large computing power~\cite{HsiangFuYuWWW2015,BinbiWSDM2014}. These techniques cannot scale to the size of Twitter with hundreds of millions of users. (In our work we do use LDA but restrict it to validation of a few thousand Twitter accounts.)

In this paper, we take a different approach to finding sensitive accounts -- one that generalizes better to unforeseen topics, is not limited by the language features, and is easily scalable to the size of Twitter.

\subsection{Our Approach}

Twitter does not enforce a real-name policy, enabling some users to adopt non-identifying pseudonyms (termed \textit{anonymous} accounts) and others to voluntarily reveal their identities by disclosing their full names (termed {\em identifiable} accounts).
A recent study leveraged Amazon Mechanical Turk (AMT) to analyze  accounts relating to sensitive topics (such as pornography, religious hatred, and drugs) and non-sensitive topics (such as news and family), finding that the sensitive accounts  have relatively large percentages of anonymous followers and relatively small percentages of identifiable followers, and vice versa for the non-sensitive accounts~\cite{PeddintiCOSN2014}.
Another study on Whisper social network has made a similar observation that anonymous content is generally more sensitive compared to non-anonymous content~\cite{correa2015whisper}.
In this paper, we make use of this observation to develop a novel data-driven approach to identify sensitive accounts. 

In this work, we consider a Twitter account to be sensitive if it has a relatively large number of anonymous followers and a relatively small number of identifiable followers. This alternative definition does not directly depend on specific sensitive words that humans choose. 
To  automatically find the sensitive accounts on Twitter, we first consider the sub-problem of automatically determining if a Twitter account is anonymous or identifiable.  
We then develop a heuristic for classifying an account as sensitive as a function of the percentages of anonymous and identifiable followers that the account has.
We applied our approach to approximately 100,000 accounts with 404 million active followers. 
In addition to detecting many of the usual suspects (accounts related to pornography, drugs, and so on), our approach uncovered many accounts related to socially stigmatized topics, such as depression, self-mutilation, obesity, and anorexia. 

It is to be noted that traditional machine learning evaluation metrics that calculate the detection performance on a pre-labeled dataset do not work in our scenario. Since content sensitivity is highly subjective, frequent disagreements among the human evaluators prevent us from utilizing manual labeling to generate the ground truth labels. This is contrary to abuse or spam account detection, where the accounts clearly violate a specific set of policies. Hence, in this work we utilize alternate validation mechanisms, such as LDA. We also check if independent human evaluators, in spite of their subjective differences, \textit{generally} agree with our account sensitivity classification results.

\subsection*{Contributions:}
\begin{itemize}
\item 
We built a machine learning classifier that can detect anonymous and identifiable Twitter accounts with more than 90\% precision. This approach makes use of name 
popularity rankings, word occurrences in Scrabble word lists (word lists  without proper nouns and names), and Twitter account profile properties. In comparison, simply checking for name occurrences from first and last name lists did not give more than 58\% precision for anonymous and 70\% precision for identifiable accounts.  
\item 
Based on linear functions of an account's percentages of anonymous and identifiable followers, we classified an account as sensitive or non-sensitive. To obtain the coefficients for the linear functions, we applied our anonymous account classifier to the followers of the known sensitive and non-sensitive accounts  studied in \cite{PeddintiCOSN2014}.
Accounts that our automated methodology labels as sensitive have high percentages of anonymous followers and  low percentages of identifiable followers. The approach does not directly define sensitivity in terms of specific words appearing in tweets, and therefore generalizes to a wider range of sensitive topics. To show that the proposed methodology can indeed process data at the scale of Twitter, we applied our methodology on  approximately 100,000 accounts having a total of 404 million active followers.   We manually analyzed 300 of the identified very sensitive accounts, and show that these accounts generalize beyond themes discovered in \cite{PeddintiCOSN2014}.
\item 
To validate our methodology, we applied Latent Dirichlet Allocation (LDA) topic analysis to  the tweets in the identified sensitive and non-sensitive accounts. The LDA analysis confirmed that there is very little overlap in the topic themes across the two groups: themes such as drugs, escort services, pornography and cybersex dominated the sensitive group; themes such as sports, weather and education are dominant in the non-sensitive group. We further validated the approach by  
asking four humans to subjectively classify 200 accounts as either sensitive or non-sensitive, and checked whether their classifications are consistent with our automated classifications. 
\end{itemize}

The rest of the paper is organized as follows.
Section~\ref{sec:ethicalconsiderations} discusses the legal and ethical considerations of this work.
Section~\ref{sec:twittersensitivebackground} provides a brief overview of Twitter, and the user categories selected. Section~\ref{sec:twittersensitivedataset} describes the collected dataset statistics. Section~\ref{sec:automatedsensitivedetection} describes our custom classifier to detect anonymous and identifiable accounts. Section~\ref{sec:sensitiveaccountdetection} describes the proposed approach to automatically detect sensitive accounts. Section~\ref{sec:twittersensitivevalidation} validates the proposed approach by performing LDA topic analysis and analyzing the detected sensitive accounts. 
Section~\ref{sec:twittersensitiverelatedwork} discusses the related work and Section~\ref{sec:conclusion} concludes the paper.

\section{Legal \& Ethical Considerations}
\label{sec:ethicalconsiderations}
To conduct this research, we programmed our crawlers to collect data from Twitter. Performing research using user data collected from online social networks can be an ethically sensitive issue. We took several precautions in our study. First, we used the official Twitter API to collect data and followed the API rate limits to ensure the load on the Twitter servers is minimized.  
Second, when collecting data using the provided Twitter API, we limited our collection to publicly available information. 
Third, we restricted our analysis to the textual account profile information alone, and gathered the profile pictures and the most recent tweets of a couple of thousand Twitter accounts 
only when absolutely necessary for machine learning feature selection and validation.

We believe that the only way to reliably estimate success rates of the proposed methodology is to conduct experiments on large real Twitter data sets. We emphasize that this research benefits Twitter and other social networks that do not enforce a real-name policy, by helping them gain visibility into how users take advantage of (or abuse) the features supporting online anonymity. Any inferences we made were based on publicly available data.

\section{Background}
\label{sec:twittersensitivebackground}

\subsection{Twitter Overview} 
When a user creates a Twitter account, he or she needs to create a \textit{profile}. The profile includes a unique alphanumeric ID (sometimes called the screen name), a name string (often containing a first and/or last name), a textual description, a profile picture, location information, and a URL (linking to other social network profiles or pointing to a website). Everything except the ID field is optional.

\textit{Tweets} are the messages posted by users. They are restricted to 140 characters and can contain text, shortened URLs, user mentions (tagging other users), and \textit{hashtags} (a metadata tag used to group messages). Each Twitter account can follow other accounts and receive their tweets. When Alice follows Bob, Bob is said to be a {\em friend} of Alice, and Alice is said to be a {\em follower} of Bob. Note that if Bob is a friend of Alice, Alice need not be a friend of Bob (unlike Facebook). Twitter provides free API access (with request limits) to obtain publicly available social network data.

\subsection{Account Categorization}
\label{sec:userclassification}
Similar to prior work~\cite{PeddintiCOSN2014}, we categorize Twitter users based on their degree of anonymity:
\begin{itemize}
\item \textbf{Anonymous} -- A Twitter account containing neither the first nor last name, and not containing a URL in the profile (which may point to a web page that identifies or partially identifies the user). 
\item \textbf{Partially Anonymous} -- A Twitter account having a first name or a last name but not both in the profile.
\item \textbf{Identifiable} -- A Twitter account containing both a first name and a last name in the profile.  
\item \textbf{Unclassifiable} -- A Twitter account that is neither Anonymous, Identifiable, nor Partially Anonymous. Accounts which have neither a first nor a last name but have a URL fall under this category. Also, Twitter accounts of an organization or a company belong here.

\end{itemize}

We recognize that Twitter does not support complete anonymity, as all users are required to choose some screen name, which is often a pseudonym. However, for all practical purposes, a pseudonym does not reveal the identity of the user. Hence we prefer to use the more commonly employed term anonymous rather than the more obscure term pseudonymous. Also, some Twitter users may choose to use fake first and last names, giving the impression they are identifiable when they are actually anonymous.  Filtering out these fake-name accounts has been shown to be highly difficult~\cite{boydFakenames2004}; it is also difficult to estimate what fraction of seemingly identifiable accounts are fake-name accounts.
As validated in Section~\ref{sec:twittersensitivevalidation}, {\em this noise does not seriously impact our ability to automatically classify accounts as sensitive or non-sensitive}. 

Twitter is plagued by unused ephemeral accounts or those created to spread spam~\cite{twitterspamcharacterization,twitterspamsuspendedaccounts}. 
To avoid any bias on the results caused by these accounts, we remove from our data sets all user accounts exhibiting signs of being ephemeral or spam. We say an account is {\em non-ephemeral} if the sum of its friends and followers is non-zero {\em and} it has posted a tweet at least six months after its creation.

Since our study revolves around using user anonymity to detect account sensitivity, 
fake accounts (like spammers) may alter our results.  Fortunately,
Twitter already puts significant effort in blocking spam.  Prior studies have shown that Twitter blocks almost 92\% of the spam accounts within 3 days of the first tweet, and all of the spam accounts (including those belonging to big spam campaigns) within 6 months~\cite{twitterspamsuspendedaccounts}. However, to be even more certain that our results are not skewed by spam accounts, we eliminated accounts that have some resemblance to reported spam account behavior, such as followers-to-friends ratio being less than 0.1. In addition to removing ephemeral and spam accounts, we also sanitized our datasets by eliminating all non-English accounts, i.e. accounts which do not report English as the language of preference in their profiles.

\section{Twitter Data Sets}
\label{sec:twittersensitivedataset}
\subsection{Labeled Training Data}
\label{sec:twittersensitivetrainingdataset}

We used supervised machine learning to automatically classify accounts as sensitive, and also classify the account followers as anonymous or identifiable. For this we leveraged Amazon Mechanical Turk (AMT) labeled data from a prior study~\cite{PeddintiCOSN2014}, which contains two distinct data sets. 

The first data set measures the prevalence of anonymity on Twitter. It contains 100,000 randomly selected accounts from a  public Twitter data set released in 2010~\cite{twittersocialmedia}. This data set was sanitized (by removing non-English, ephemeral, and spam accounts)
and labeled using AMT. The second dataset studies the influence of content sensitivity on user anonymity. It was created by picking 47 Twitter accounts related to the different sensitive categories (such as pornography, escort services, and racism), and 20 accounts related to non-sensitive categories (such as news sites, family recreation, and movies). The followers of these 67 accounts were sanitized 
and labeled using AMT. The combined labeled accounts from the two data sets constitute our training set. The distribution of the accounts across the different anonymity categories is shown in Table~\ref{tab:trainingset}.

\begin{table}[thbp]
\small
\begin{center}
\caption{Training Set for Machine Learning}
\label{tab:trainingset}
\begin{tabular}{| l | c |}
\hline
\textbf{Label} & \textbf{\# of Twitter Accounts}\\
\hline
Identifiable & 66,903 (51.3\%)\\
\hline
Partially Anonymous & 27,734 (21.2\%)\\
\hline
Anonymous & 19,890 (15.2\%)\\
\hline
Unclassifiable & 16,105 (12.3\%) \\
\hline
\hline
Total & 130,632 \\
\hline  
\end{tabular}
\end{center}
\vspace{-1.0em}
\end{table}

\subsection{Experimental Data Set}
\label{sec:twittersensitivetestdata}
To test if we can detect sensitive Twitter accounts by analyzing their follower anonymity characteristics, and to evaluate if our approach can process data on the Twitter scale, we need a new data set that is sufficiently large. We crawled Twitter from May 31 - Aug 7, 2014. Starting from the 67 hand-picked accounts in the training set belonging to the sensitive and non-sensitive topics (the seed list), we crawled outwards -- crawling two levels down of account followers but randomly selecting branches and accounts.
We sanitized the resulting set by removing all non-English accounts (as we discuss later, in spite of this filtering we identify some accounts that post non-English tweets) and accounts with $<$200 \textit{active} (non-ephemeral and non-spam) followers. Our resulting data set has 93,042 accounts with approximately 404 million active followers. We applied our sensitive account discovery methodology to this data set.

\section{Automating Identification of Anonymous Accounts}
\label{sec:automatedsensitivedetection}

It has been shown that accounts most people would deem as sensitive typically have a relatively large number of anonymous followers and a relatively small number of identifiable followers~\cite{PeddintiCOSN2014}. In this paper we consider automatically identifying sensitive accounts based on these characteristics.
To this end, we need an automated procedure for determining, for each of the account's followers,  whether  the follower is anonymous or identifiable (or neither).  

This section first explores a simple technique: using first and last name lists to categorize an account as identifiable or anonymous.  We then discuss how machine learning allows us to classify accounts with a much higher degree of precision and recall.

\subsection{Categorization Using Name Lists}
\label{sec:namelistmembership}

Since the definitions for all the user groups (Section~\ref{sec:userclassification}) rely on the presence/absence of first/last names in the Twitter account's profile, a straightforward approach to classify accounts would be to check for memberships in existing lists of common first and last names. To this end, we obtained public name lists from the United States Census~\cite{censusnamesdatabase} and 
United States Social Security Administration~\cite{ssanamesdatabase}. We gathered a total of 91,340 first names and 165,640 last names.
We observed that simply checking for occurrences in the name lists results in very poor anonymous and identifiable detection rates on the training data set. The detection performance is shown in Table~\ref{tab:namelistperformance}. 

\begin{table}[thbp]
\small
\begin{center}
\caption{Detection Performance when Simply Checking Name Lists}
\label{tab:namelistperformance}
\begin{tabular}{| l | c | c |}
\hline
\textbf{Label} & \textbf{Precision} & \textbf{Recall}\\
\hline
Anonymous & 0.58  & 0.35\\
\hline
Identifiable & 0.70 & 0.83\\
\hline  
\end{tabular}
\end{center}
%\vspace{-1.0em}
\end{table}

We noticed the occurrence of common English words in the first and last name lists -- such as Gay, Love, Clay, Crystal, May -- was one of the primary reasons for the low detection performance.
As most of the Twitter names contain phrases, parts-of-speech taggers were not very helpful in differentiating the two usages. 
Even after imposing structural constraints for the name string (such as following \textit{FirstName MiddleName LastName}, \textit{FirstName MiddleInitial LastName}, or \textit{FirstName LastName} formats with appropriate memberships in first and last name lists) we did not obtain significant improvement in the performance.  

\subsection{Feature Selection for Classification}
\label{sec:featureselection}

Even though membership in name lists did not work well, this does not mean that accounts cannot be effectively classified by other means.  Other profile properties, such as the rank of occurring names in the first and last name lists, or the occurrence of non-names in the name field can also be examined. 
We used a machine learning classifier as it is capable of accommodating all these different features, and considered the implementations available in the Weka toolkit~\cite{weka}.
We initially considered all the features available from a Twitter account's profile, and later refined our selection. We mention which features and representations have the most impact.

It has been observed that pictures sometimes present an effective way to identify users.  To estimate the usefulness of Twitter profile pictures in deducing the identity of an account, we conducted a small study. We obtained the profile pictures for 1,000 random accounts from the training set in each of the anonymous and identifiable categories. For the identifiable accounts,
61.5\% of profile pictures contained discernible faces.
Of the 1,000 anonymous profile pictures, only 119 had pictures with discernible human faces. We conducted a Google image search~\footnote{\url{http://images.google.com/}} on all the 119 pictures and observed the following:
(i) 24 pictures belonged to popular celebrities/icons; 
(ii) for 13 pictures the image search was able to point out a name, because the same picture was used on 
personal web pages or Facebook/Google+ profiles;
(iii) in the remaining 82 cases, image search did not return any results (except pointing out other pictures with similar color variations/structure). 
As the profiles pictures help in deducing identities of just 1.3\% of anonymous accounts, and the identification procedure is very cumbersome to automate when there are millions of accounts, we did not include profile pictures in the feature selection process.  

We extracted other pieces of naming related information available from the screen name and the name fields.
We checked for the structural constraints on the name, and take into account the popularity ranks of the occurring names in the name lists. As users often combine their first and last names into a single word (like `Adam J Smith' occurring in the word `adamjsmith'), we considered first or last names occurring as a sub-string. To limit classification errors due to English words occurring in name lists, we leveraged Scrabble word dictionaries (English word lists used in the Scrabble board game, generally do not contain proper nouns)~\footnote{\url{http://www.freescrabbledictionary.com/sowpods.txt}}~\footnote{\url{http://www.isc.ro/en/commands/lists.html}}
and word frequencies obtained from the British National Corpus~\footnote{Adam Kilgarriff, BNC database and word frequency lists, \url{http://www.kilgarriff.co.uk/bnc-readme.html}}.

In addition to these name related features, we extracted other information available from a Twitter account's profile. We considered the number of friends, followers, tweets, and \textit{favorited} tweets (favoriting is done to express likeness or for archiving the tweet). Twitter users can be grouped into \textit{lists} for easy reading of tweet updates from the group members. We considered the number of lists an account has membership in. Twitter provides a \textit{protected} privacy feature to hide the activity (such as tweets, friends, and followers) from being publicly visible. Also, a Twitter user can associate his geographical information to his Twitter account by activating the \textit{geo-tagging} feature. We checked if a Twitter profile uses the protected and geo-tagging functionalities. 

After testing various configurations with the features and feature representations, we chose 16 features: 12 numeric and 4 boolean. The features are listed in Table~\ref{tab:features}. In case the profile does not contain any first or last name occurring in the name lists, then we considered the maximum possible integer for the corresponding name popularity rank features. Similarly if the detected first or last name does not occur in the Scrabble word list, or if no first and last names are detected, then the Scrabble word frequency ranks are initialized to the maximum possible integer. Table~\ref{tab:features} also indicates the information gain values for each feature. A higher gain value indicates a stronger influence in detecting anonymous and identifiable accounts. The top 3 features that help detect anonymous or identifiable accounts are highlighted in bold.

\begin{table*}[thb]
\small
\begin{center}
\caption{Selected Feature Set for Machine Learning Classification}
\label{tab:features}
\begin{tabular}{| l | p{7cm}| c | c |}
\hline
\multirow{2}{*}{\textbf{Type}} & \multirow{2}{*}{\textbf{Feature}} & \multicolumn{2}{c|}{\textbf{Information Gain}} \\
\cline{3-4}
 &  & \textbf{Anonymous} & \textbf{Identifiable} \\
\hline
\multirow{12}{*}{Numeric} & \# of friends & 0.015026 & 0.04283\\
\cline{2-4}
 & \# of followers  & 0.009373 & 0.04705\\
\cline{2-4}
 & followers-to-friends ratio & 0.000822 & 0.00468\\
\cline{2-4}
 & \# of user list memberships & 0.002298 & 0.02168\\
\cline{2-4}
 & \# of tweets & 0.005205 & 0.02584\\
\cline{2-4}
 & \# of favorite tweets & 0.016624 & 0.0166\\
\cline{2-4}
 & number of parts/words in the name string & 0.033509 & \textbf{0.19704}\\
\cline{2-4}
 & popularity rank of occurring first name in the name list & \textbf{0.156569} & \textbf{0.23648} \\
\cline{2-4}
 & popularity rank of occurring last name in the name list & \textbf{0.062874} & \textbf{0.2413}\\
\cline{2-4}
 & \# of Scrabble words present in the name & 0.023888 & 0.06242\\
\cline{2-4}
 & word frequency rank of occurring first name in the Scrabble list & 0.040174 & 0.06794\\
\cline{2-4}
 & word frequency rank of occurring last name in Scrabble list & 0.033384 & 0.07316\\
\hline
\hline 
\multirow{4}{*}{Boolean} & enabled \textit{protected} privacy feature & 0.00023 & 0.00129\\
\cline{2-4}
& enabled geo-tagging for tweets & 0.002429 & 0.00105\\
\cline{2-4}
& includes a url in the profile & \textbf{0.091024} & 0.00821\\
\cline{2-4}
& name follows structural constraints & 0.044985 & 0.185\\
\hline
\end{tabular}
\end{center}
\vspace{-1.5em}
\end{table*}

\subsection{Customized Classifier for Account Anonymity Classification}
\label{sec:customclassifier}
As Table~\ref{tab:trainingset} indicates, we have more than 2 classes. When there are multiple classes (with dissimilar size distributions), machine learning classifiers by default try to optimize the overall achievable accuracy at the cost of lower precision and recall for small sized classes. In our case, when using all four classes, we noticed low precision and recall values being reported for anonymous accounts. Even using different classifiers,
optimizing classifier parameters, and using meta-classifiers (such as multi-class classification or cost-sensitive classification) did not improve the performance. 
(In multi-class classification, multiple classifiers are built, one for each pair of classes, and the final classification label for an instance is determined based on a voting mechanism. In cost-sensitive classification, additional cost for misclassifying accounts as anonymous or identifiable is imposed. These costs are used to re-weight the training instances or to predict the class with minimum expected misclassification cost.)

\begin{figure}[thbp]
\begin{center}
  \centering
    \caption{Machine Learning Training}
  \label{fig:training}
     \includegraphics[width=0.45\textwidth]{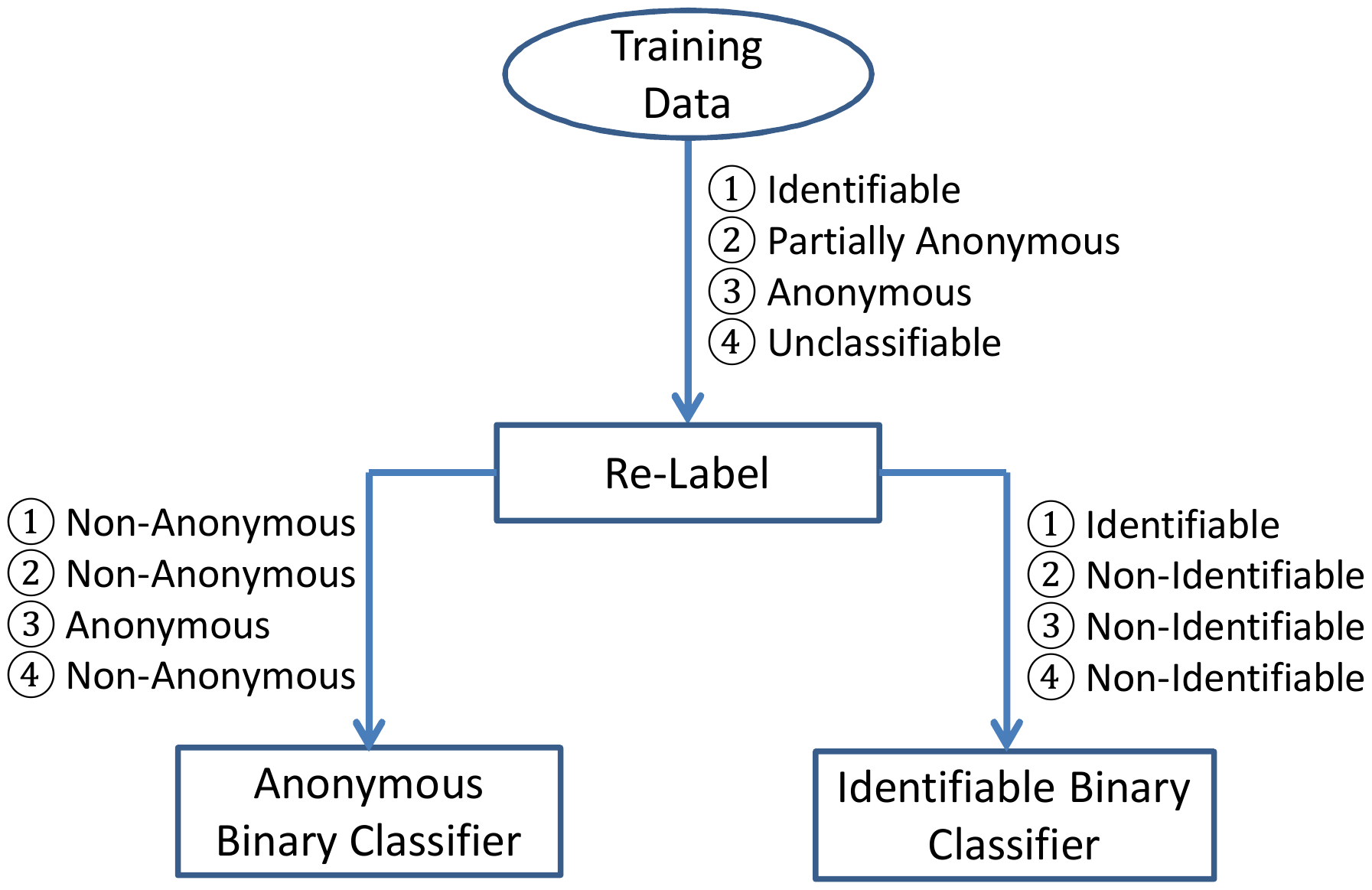}

\end{center}
%\vspace{-1.5em}
\end{figure}

We developed a classifier that converts the four-class classification problem into two binary classification problems: one that classifies each account either as anonymous or non-anonymous; the other that classifies each account as either identifiable or non-identifiable. The results of the two classifiers are then combined to classify each account as ``anonymous,'' ``identifiable'' or ``unknown'' as described below. 

The training phase is shown in Figure~\ref{fig:training}. The training data containing four classes gets relabeled into two data sets containing the same number of training instances as the original. In the first data set, all the instances for classes other than the anonymous class get re-labeled as `Non-Anonymous,' and this data set is passed to a binary classifier that is optimized for detecting anonymous accounts. In the second data set, all the instances for classes other than the identifiable get re-labeled as `Non-Identifiable,' and this data set is passed to a binary classifier optimized for detecting identifiable accounts. 
Both the binary classifiers use Random Forest with 100 trees as the base classifier. The choice of the classifier and the number of trees was based on the cross-validation performance and out-of-bag error~\cite{introductiontostatisticallearning}. These classifiers are also cost-sensitive meta classifiers, where higher costs are imposed for mis-classifying instances as anonymous/identifiable. 

\begin{figure}[thbp]
\begin{center}
  \centering
    \caption{Anonymous and Identifiable: Precision vs. Recall Trade-off}
  \label{fig:precisionrecall}
     \includegraphics[width=0.45\textwidth]{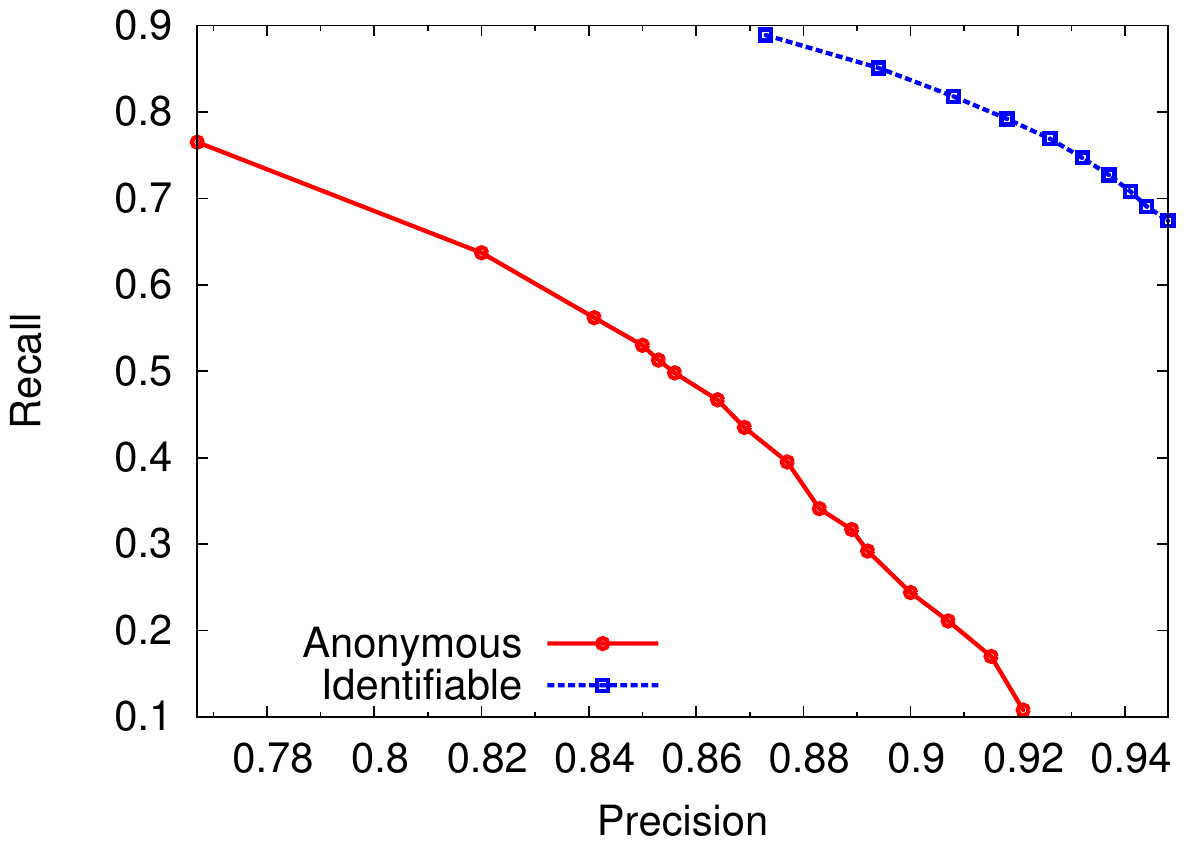}
\end{center}
%\vspace{-1.0em}
\end{figure}

The mis-classification costs are chosen independently for the two binary classifiers.
By varying the cost parameters for the two binary classifiers, we can trade-off the precision and recall values. 
For both the classes, the precision and recall trade-off values based on 10-fold cross validation are shown in Figure~\ref{fig:precisionrecall}. 
For this study we chose the cost parameters such that at least 90\% precision is possible for both anonymous and identifiable classes. Table~\ref{tab:precisionrecall} shows the choice of cost parameters, and the resulting precision and recall values. 
Although the precision for both anonymous and identifiable accounts is high, and the recall for identifiable accounts is high, the recall for anonymous accounts is 24.4\%.
Unlike spam or abuse detection techniques, our sensitive account detector does not necessitate identifying all of the anonymous or identifiable accounts -- {\em What is most important is identifying a significant fraction with low error rates}. Hence, high recall values are not absolutely necessary, and our results in the subsequent sections validate this.

\begin{table}[thbp]
\small
\begin{center}
\caption{Cost Parameters and Classifier Performance Based on 10-fold Cross Validation}
\label{tab:precisionrecall}
\begin{tabular}{| l | c | c | c |}
\hline
\textbf{Label} & \textbf{Cost Parameter} & \textbf{Precision} & \textbf{Recall}\\
\hline
Anonymous & 9.5 & 0.90  & 0.244\\
\hline
Identifiable & 6.0 & 0.932 & 0.747\\
\hline  
\end{tabular}
\end{center}
%\vspace{-1.75em}
\end{table}

\begin{figure}[thbp]
\begin{center}
  \centering
  \caption{Machine Learning Testing}
  \label{fig:testing}
     \includegraphics[width=0.45\textwidth]{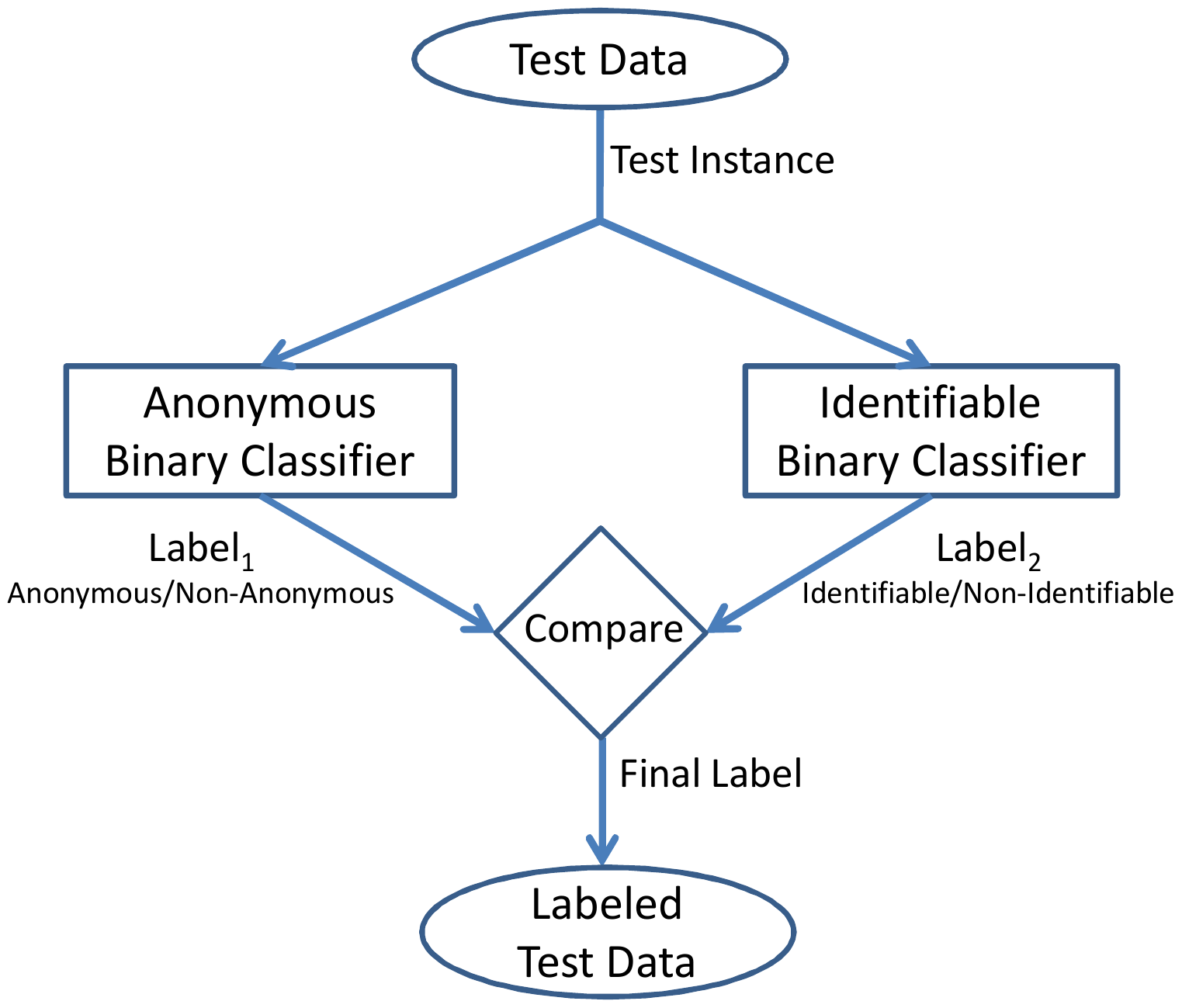}
\end{center}
%\vspace{-2.75em}
\end{figure}

The testing phase is shown in Figure~\ref{fig:testing}. Each test instance gets passed to each of the binary classifiers, which independently assigns a label to the instance. We determine the final label based on the decision table in Table~\ref{tab:labeldecision}. ``Unknown'' means we do not attempt to classify the account. After classifying approximately 404 million followers in the test data, we did not come across any case where an account was classified as both anonymous and identifiable by the binary classifiers (corresponding to the fourth row in Table~\ref{tab:labeldecision}).

\begin{table}[thbp]
\small
\begin{center}
\caption{Deciding Final Label for a Test Instance}
\label{tab:labeldecision}
\begin{tabular}{| c | c | c |}
\hline
\textbf{Label$_{1}$} & \textbf{Label$_{2}$} & \textbf{Final Label}\\
\hline
Anonymous & Non-Identifiable & Anonymous \\
\hline
Non-Anonymous & Identifiable & Identifiable \\
\hline
Non-Anonymous & Non-Identifiable & Unknown \\
\hline
\multirow{2}{*}{Anonymous} & \multirow{2}{*}{Identifiable} & Unknown \\
& & \textit{(Did not occur)} \\
\hline  
\end{tabular}
\end{center}
%\vspace{-1.0em}
\end{table}

\section{Sensitive Account Discovery}
\label{sec:sensitiveaccountdetection}

The previous subsection showed how we can classify many Twitter accounts with high precision as anonymous or identifiable. We refer to these accounts
as ``discovered  anonymous'' and ``discovered identifiable'' accounts. 
We use these to find sensitive accounts.

As mentioned earlier in Section~\ref{sec:introduction}, in our approach we suspect that an arbitrary account is sensitive (non-sensitive) if it has a relatively large (small) number of anonymous followers and a relatively small (large) number of identifiable users. 
To this end, we used the percentages of ``discovered  anonymous'' and ``discovered identifiable'' accounts as proxies for the actual percentages of  anonymous and identifiable accounts, since we can readily obtain the percentages of the discovered accounts automatically using machine learning, as shown in  Section~\ref{sec:customclassifier}.

\begin{figure}[thb]
\begin{center}
  \centering
    \caption{Scatter Plot of Sensitive and Non-Sensitive Accounts Based on Discovered Anonymous and Identifiable Followers}
  \label{fig:svmscatterdiagram}
     \includegraphics[width=0.45\textwidth]{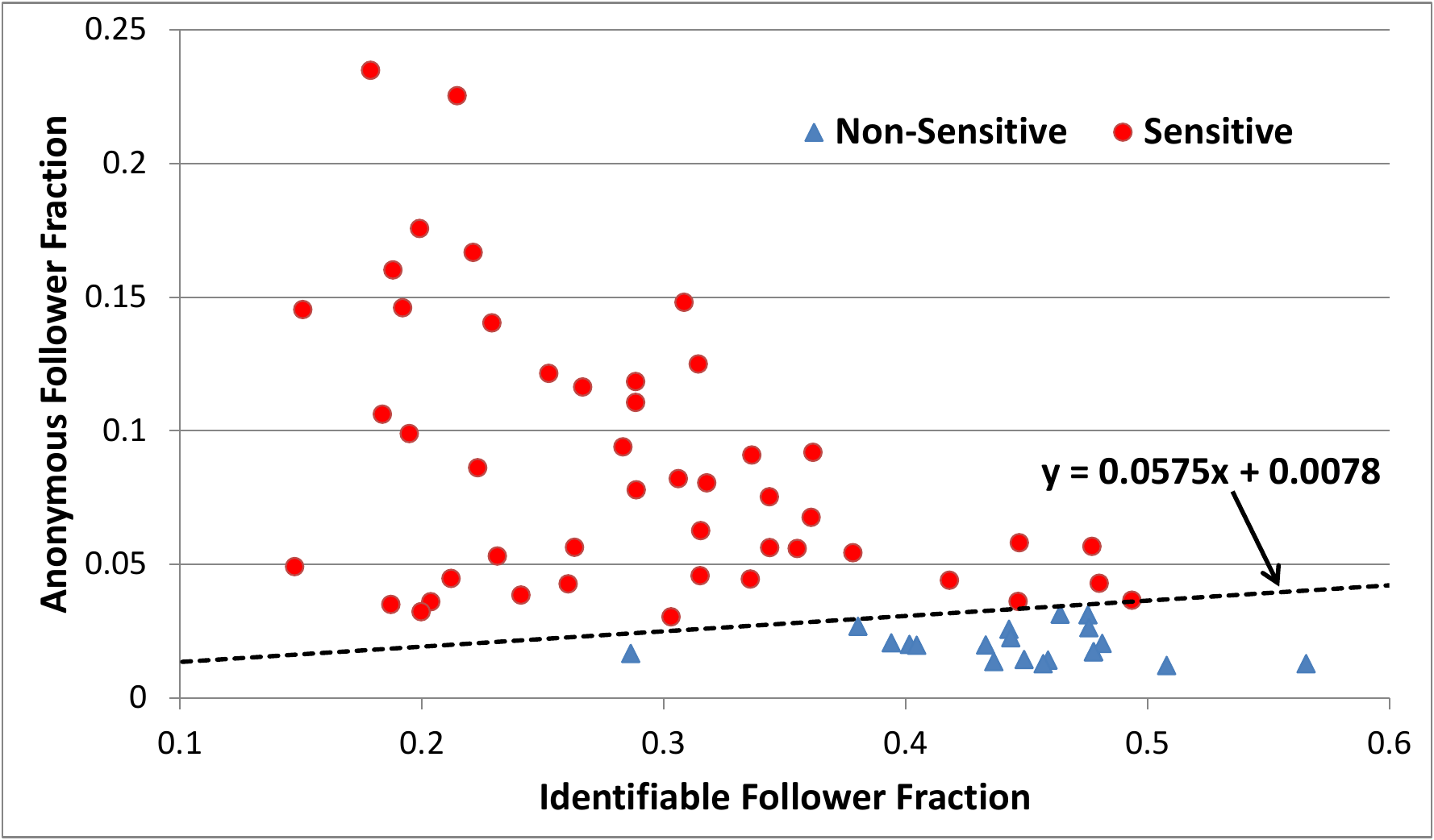}
%\vspace{-1.5em}
\end{center}
\end{figure}

To precisely state what we mean by large and small percentages of discovered anonymous and identifiable accounts, we returned to the 67 manually chosen sensitive and non-sensitive accounts  from the recent work~\cite{PeddintiCOSN2014}. 
Using the automated methodology developed in Section~\ref{sec:customclassifier}, we classified the followers of all these 67 accounts into Anonymous, Identifiable and Unknown, and determined the fractions of discovered anonymous and identifiable followers for each account. Figure~\ref{fig:svmscatterdiagram} shows a scatter diagram, where each circle (triangle) corresponds to one of the chosen sensitive (non-sensitive) accounts. For each account, Figure~\ref{fig:svmscatterdiagram} shows the fractions of discovered anonymous and identifiable followers. Strikingly, the sensitive accounts all lie at the top-left, and the non-sensitive accounts all lie at the bottom-right of the plot. Using a Support Vector Machine (SVM) classifier, we can  separate the sensitive and non-sensitive accounts in the scatter diagram. The SVM classifier uses a linear kernel, with the regularization parameter $C$ as 5,000 (resulting in a narrow-margin linear hyperplane).  The linear hyperplane equation obtained is $y=0.0575x+0.0078$. 

It is to be noted that sensitivity is not a binary concept, but rather a nuanced measure that should we viewed on a continuum~\cite{Peddinti2014,correa2015whisper}.
We say that we {\em suspect a Twitter account to be sensitive} if  $y > 0.0575x + 0.0078$, where $y$ is the fraction of 
discovered anonymous followers the account has and $x$ is the fraction of discovered identifiable followers the account has. Further, if $y >> 0.0575x + 0.0078$, we suspect the account to be {\bf very sensitive}. In a similar manner, we suspect accounts to be non-sensitive and very non-sensitive by reversing the inequalities. While there might not be perceptible difference between the accounts close to the linear hyperplane, the extremes should exhibit clearly distinct behavior. For a given account, the $x$ and $y$ values
are determined by the automatic classification technique described in Section~\ref{sec:customclassifier}. Note that although the fraction of identifiable followers plays a role in our sensitive account detection methodology, it is actually the fraction of anonymous followers that carries the largest weight.  

When we applied this methodology to the 93,042 random test accounts, 59.3\% of the accounts lie on the sensitive side of the linear hyperplane, and 40.7\% lie on the non-sensitive side. Admittedly, this approach also has some subjectivity in the definition, as the coefficients in the separating line depend
on the 67 hand-chosen sensitive and non-sensitive accounts.  If we were to choose a different set of accounts, the line $y = 0.0575x + 0.0078$ would change some. However, we note that this approach does not directly depend on pre-selected sensitive keywords and, as we will show,  identifies many sensitive categories that were not covered by our hand-chosen 67 accounts (and is hence generalizable). Also, this approach has an additional advantage that it is not dependent on language features and can be easily extended to content across a wide variety of languages. To minimize the impact of small changes in the coefficients of the line, 
we henceforth only consider the accounts that are far away from the linear hyperplane (i.e., the very sensitive and very non-sensitive accounts).

\section{Validation}
\label{sec:twittersensitivevalidation}

To validate that our methodology can indeed find sensitive accounts on Twitter, we take three approaches. First, we use Latent Dirichlet Allocation (LDA) to show that the topics being discussed in the very sensitive accounts identified by our methodology are indeed immensely different from those being discussed in the very non-sensitive accounts. Second, we manually look into the individual accounts classified as very sensitive and see if they are indeed about topics that many would consider sensitive/controversial. Third,
we ask four humans to subjectively classify 200 accounts as either sensitive or non-sensitive, and then check whether their classifications are consistent with our automated classifications.

\subsection{LDA Topic Analysis}
\label{sec:ldatopicanalysis}

Using LDA, we analyzed the tweets to determine if the sensitive and non-sensitive accounts are indeed talking about different topics.
In LDA, each document is a collection of words; each topic is considered to be a probabilistic distribution over a collection of words; and each document is considered to be a probabilistic distribution over the topics. The goal of LDA is to study the \textit{observable} word occurrences in the documents and to determine the \textit{latent} topics in the document collection, as well as the topics discussed by each document~\cite{blei2003lda}.

We picked the 1,000 sensitive and 1,000 non-sensitive accounts that are farthest from the linear SVM hyperplane,
and obtained their most recent 200 tweets. The median number of tweets for these 2,000 accounts is 196. 
In our context, there is one document for each of the 2,000 accounts, and a document is all of the 200 tweets published by the account. 
We performed the LDA topic analysis on these 2,000 documents using the Stanford Topic Modeling Toolkit~\footnote{\url{http://nlp.stanford.edu/software/tmt/tmt-0.4}}. Specifically, we used the CVB0LDA model, which is the Collapsed Variational Bayes approximation to the LDA.

When performing LDA, we need to choose certain parameters -- the number of topics and the Dirichlet prior probabilities for the topic-word and document-topic distributions.
To choose the number of topics, we performed a \textit{Perplexity} analysis~\cite{blei2003lda}. For this, the document corpus is split into two subsets (80-20 ratio): one used for training the LDA model, and the other used for evaluating the generated model on unseen data and generating a score called perplexity. 
For each choice of the number of topics, we calculated the perplexities. A lower value for the perplexity indicates a better model~\cite{blei2003lda}. 
Based on this analysis we chose 250 topics for our study.
After testing some commonly used configurations for prior probabilities for topic-word and document-topic distributions, based on performance we chose Symmetric Dirichlet priors with 0.01 probability for both. In summary, we performed the LDA analysis on the 2,000 documents using the CVB0LDA model with 250 topics and Symmetric Dirichlet priors with 0.01 probability as the parameters.

\begin{figure*}[thb]
\begin{center}
  \centering
  \caption{Topic Usage Across Sensitive and Non-Sensitive Accounts Ordered by the Decreasing Ratio of Cumulative Weight of Sensitive Over Non-sensitive (Numbers on x-axis do not Correspond to Topic Numbers)}
  \label{fig:ldatopicusage}
     \includegraphics[width=1.0\textwidth]{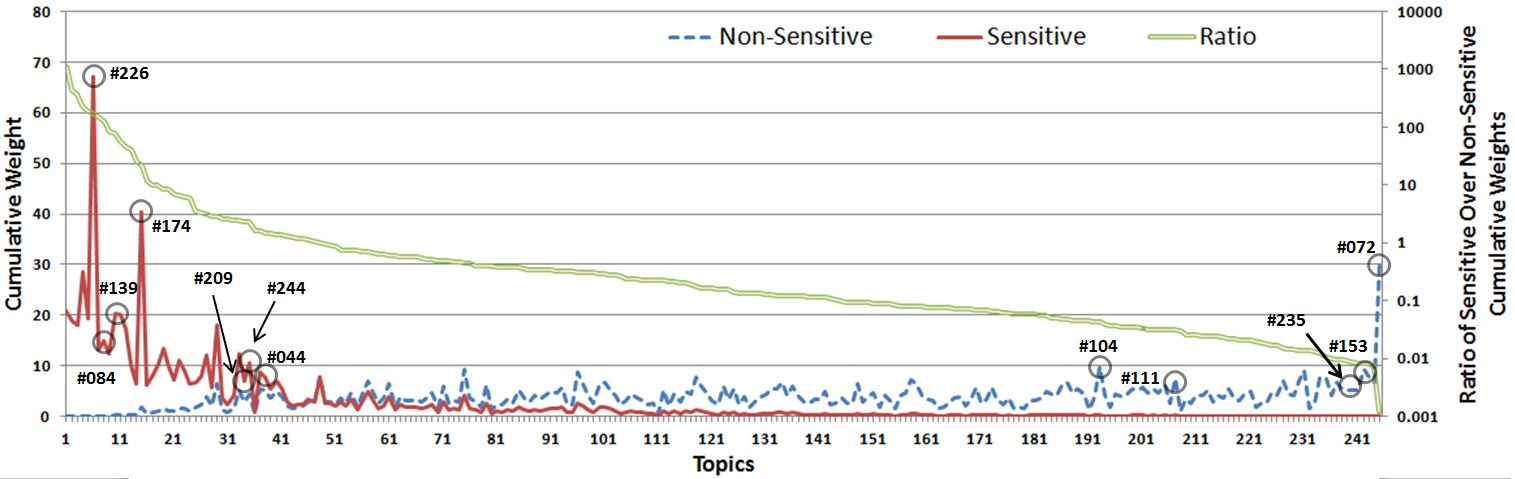}
%\vspace{-2.0em}
\end{center}
\end{figure*}

LDA gives us the topic weight distribution for each document/account. So for each identified topic, we then calculate the cumulative weight for that topic across all the sensitive accounts, and also do the same for the non-sensitive accounts. (Note that the total cumulative weight to be distributed across all the topics is the same for the sensitive and non-sensitive accounts, as there are 1000 accounts
of each.) Therefore, each of the 250 topics has  two cumulative weights, one for the accounts in the sensitive category and one for the non-sensitive category; moreover, if a topic has higher cumulative weight for the sensitive category, then we can say it has a stronger affinity with the sensitive accounts. The distribution of cumulative weights for topics across sensitive and non-sensitive categories is shown in Figure~\ref{fig:ldatopicusage}, where each point on the x-axis indicates one of the 250 topics. The topics are ordered by decreasing ratio of the cumulative weight of sensitive over non-sensitive categories. (The numbering on the x-axis does not correspond to the topic numbers.) Figure~\ref{fig:ldatopicusage} clearly shows that certain topics are very popular among the sensitive accounts, but have little or no interest to {\em all}
the non-sensitive accounts.

Some of the topics with the highest cumulative weights from the sensitive and non-sensitive accounts, and the top terms occurring in each topic, are shown in Table~\ref{tab:ldatopterms} (ratio rank is the position of the topic when ordered according to the decreasing ratio of cumulative weights of sensitive over non-sensitive accounts). These topics are also highlighted in Figure~\ref{fig:ldatopicusage}. We purposefully selectively list some of the top topics, rather than list the top-k topics, as per the cumulative weight ratio, because we want to describe the peaks in Figure~\ref{fig:ldatopicusage} and also highlight the theme diversity, as there were numerous topics belonging to the same general theme. For example, among the top 15 topics dominant among the sensitive accounts, 12 belonged to pornography, 2 belonged to cybersex, and 1 belonged to drugs/marijuana themes.

We clearly see the distinction between the topic themes in the sensitive and non-sensitive categories, giving credence to our claim that the sensitive accounts are indeed talking about different topics than the non-sensitive accounts. 
In Figure~\ref{fig:ldatopicusage}, most of the peaks in the sensitive category belong to sex-related topics. For example, topic \#174 relates to adult swingers (containing words related to sharing nude pictures or inviting sexual partners). But we also see many other non-sex-related topics, such as pregnancy, or sharing personal feelings or experiences. Though not listed in Table~\ref{tab:ldatopterms}, we noticed the presence of other topics -- such as weight loss (related to both regular and extreme cases of anorexia), fitness, marriage problems, and relationships -- receiving high cumulative weight across the sensitive accounts. Comparatively, all the topic themes dominating the non-sensitive accounts relate to ``mundane topics'' such as coupons, weather, sports, education, and so on.

\begin{table*}[tbh]
\small
\begin{center}
\caption{The Diversity of Top Topics in Sensitive and Non-Sensitive Categories (Pornographic words are in asterisks)}
\label{tab:ldatopterms}
\begin{tabular}{| c |  c | c | c | p{4.5cm} | p{1.5cm} |}
\hline
\multirow{3}{*}{\textbf{Topic ID}} &\textbf{Ratio} & \multicolumn{2}{|c|}{\textbf{Cumulative Weight}} & \multirow{3}{*}{\textbf{ Top Terms}} & \multirow{3}{*}{\textbf{Theme}}\\
\cline{3-4}
 & \textbf{Rank} & \textbf{Sensitive} & \textbf{Non-} & & \\
 & & & \textbf{Sensitive} & & \\
\hline 
226 & 6 & 67.26 & 0.39 & 
p***y, c**k, f**k, c*m, a**, hard, h***y, s**y, wet, hot, wants, f**king, l**k, s**k, nice
& Pornography\\
\hline 
084 & 8 & 14.94 & 0.12 & 
kik, h***y, s*x, p***y, pics, skype, couples, retweet, guys, m**f, wants, add, chat, a**e, kikme
& Cybersex\\
\hline
139 & 11 & 20.03 & 0.34 & 
weed, smoke, high, smoking, f**k, blunt, bong, shit, stoned, stoner, hit, stonernation, bowl, stoners, joint
& Drugs - Marijuana\\
\hline
209 & 34 & 7.05 & 2.98 &  
skinny, thinspo, anorexia, fat, eat, depression, ana, eating, hungry, look, stomach, dont, true, feel, fucking
& Anorexia\\
\hline  
244 & 35 & 10.63 & 4.53 & 
feel, line, friends, care, friend, cross, things, person, miss, beautiful, matter, say, cause, thing, afraid
& Feelings\\
\hline 
 044 & 38 & 7.68 & 5.19 & 
baby, pregnant, really, congrats, married, girl, congratulations, beautiful, little, mom, gonna, daughter, getting, guys, amazing
 & Pregnancy\\
\hline
\hline
072 & 250 & 0.04 & 31.18 & 
tomorrow, team, awesome, big, excited, congrats, family, check, game, ready, fun, year, looking, let, proud
& Generic\\
\hline
153 & 247 & 0.08 & 9.28 & 
school, students, iaedchat, teachers, learning, join, year, edchat, classroom, team, edtech, teacher, satchat, kids, educators
& Education\\
\hline
235 & 244 & 0.05 & 5.22 & 
coupon, matches, coupons, store, tesco, team, cvs, deals, walgreens, dixie, free, stores, meridian, deal, aid
& Deals / Coupons\\
\hline
111 & 207 & 0.23 & 7.37 & 
warning, severe, county, storms, thunderstorm, rain, radar, afternoon, temps, evening, mph, tornado, winds, moving, issued
& Weather\\
\hline
104 & 193 & 0.41 & 9.69 & 
mate, game, arsenal, man, pal, lads, goal, cup, cheers, big, league, season, city, liverpool, world
& Sports\\
\hline
\end{tabular}
\end{center}
%\vspace{-1.0em}
\end{table*}

\subsubsection{Topic Overlap}
Figure~\ref{fig:ldatopicusage} shows the ratio of cumulative weights for sensitive and non-sensitive accounts across 250 topics. There are just 34 topics (out of 250) for which the ratio is in the range of [0.5,2]. But perhaps most collections of two sets of 1,000 accounts would have similar topic properties?

\begin{figure}[thb]
\begin{center}
  \centering
  \caption{Comparing Ratios of Cumulative Topic Weights Across Different Account Groups}
  \label{fig:ratiocomparison}
     \includegraphics[width=0.45\textwidth]{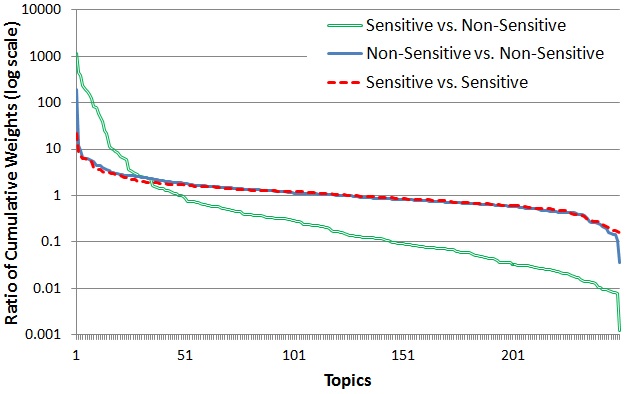}
%\vspace{-1.5em}
\end{center}
\end{figure}

To explore this issue more deeply, we conducted LDA topic analysis on additional sets of accounts. The first set contained two sensitive groups, each containing 1,000 accounts picked randomly from the farthest 10,000 very sensitive accounts from the SVM hyperplane. The second set contained two non-sensitive groups, each with 1,000 accounts picked randomly from the farthest 10,000 very non-sensitive accounts. We performed LDA topic analysis on each of the two sets independently (using the same LDA parameters as before), and determined the topic usage across the different accounts. For each set, we determined the cumulative weight for each identified topic across the two groups of accounts.  The cumulative weight ratio distributions for the identified topics in each set are shown in Figure~\ref{fig:ratiocomparison}. The results for the first set are represented as \textit{Sensitive vs. Sensitive}, and those for the second set are represented as \textit{Non-Sensitive vs. Non-Sensitive}. For comparison, we also include the ratio graph in Figure~\ref{fig:ldatopicusage} as \textit{Sensitive vs. Non-Sensitive}. Notice that the Sensitive vs. Sensitive ratio curve and the Non-Sensitive vs. Non-Sensitive ratio curves are very flat. This indicates that the two groups of sensitive accounts are largely talking about similar things; similarly, the two groups of non-sensitive accounts are largely talking about similar things. Interestingly, the ratio distributions for Sensitive vs. Sensitive and Non-Sensitive vs. Non-Sensitive are nearly identical. However, the curve comparing sensitive accounts with non-sensitive accounts exhibits a distinctly stronger decreasing trend over a much wider range of ratios, confirming that the sensitive accounts and non-sensitive accounts are indeed talking about different topics.

\subsection{Types of Very Sensitive Accounts}
\label{sec:sensitiveaccounttypes}

To further evaluate whether our methodology is indeed identifying sensitive accounts, we manually inspected the top 300 very sensitive accounts as determined by our methodology, and assigned each account to a category. Table~\ref{tab:sensitiveaccounts} lists the number of accounts that fall into each category. The miscellaneous accounts belong to profiles of individuals (almost all self-identify as females), or ones that share multimedia, have non-English tweets, and are protected or de-activated (after our initial data gathering, some accounts were de-activated). 

As expected, pornography, drugs, and adult content are pervasive in the sensitive accounts. There were also numerous  swinger accounts involving married couples sharing intimate pictures of themselves and inviting swingers. 
Most of the pictures uploaded by accounts belonging to these sex-related categories contain nudity, but they are not tagged as such, violating Twitter's media policies~\cite{twittermediapolicy}. 

We identified several accounts discussing drugs, such as marijuana. These users often share their experiences or invite others for an in-person meeting. Some of these accounts actually claim to sell or deliver marijuana. With marijuana being illegal in many countries and states, local authorities could be potentially interested in using our methodology for tracking these accounts and flagging them for take down. 

Though Table~\ref{tab:sensitiveaccounts} lists just one account supporting and fighting for rights of lesbian, gay, bisexual and transgender groups, we observed many accounts belonging to this group among the top 1000 sensitive accounts. For many people, disclosing one's sexual orientation is a sensitive issue and hence users may prefer not to identify themselves when following these accounts. Interestingly, we observed many anonymous followers for Twitter accounts self-identifying as high school and college females. 
Accounts self-identifying as female fitness and yoga accounts, in which users discuss their diet and upload pictures of their progress, also had many anonymous followers. 

In addition to these themes, we discovered several accounts that deal with severe cases of anorexia, social anxiety, depression and suicidal tendencies. 
The existence of accounts related to stigmatized topic themes in the very sensitive category testifies to the importance of privacy and anonymity in our society. 
People would be much less inclined to use these \textit{self-help} accounts without anonymity. In fact, we noticed some of these accounts uploading pictures after having physically abused their bodies. 
Health institutions are already employing Twitter for real-time monitoring of individuals with suicidal tendencies~\cite{twittersuicidetracking}. They could potentially use our methodology to identify people in urgent need of support.

\begin{table}[thb]
\small
\begin{center}
\caption{Top Sensitive Accounts and their Themes}
\label{tab:sensitiveaccounts}
\begin{tabular}{| p{1.5cm}| p{6cm} |}
\hline
\textbf{\# of} & \multirow{3}{*}{\textbf{Theme}}\\
\textbf{Sensitive} & \\
\textbf{Accounts} & \\
\hline
106 & Couples sharing their sexual escapades or looking for swingers. Pictures are not tagged for nudity.\\
\hline
47 & Arabic Adult/Gay Content. Pictures are not tagged for nudity.\\
 \hline
 21 & Relate to pornography/adult content. Pictures are not tagged for nudity. \\
 \hline
 12 & Related to drugs, such as marijuana. Some accounts offer drug delivery services. \\
 \hline
 9 & Accounts self-identifying as high school and college girls (some in their teens).\\
 \hline
 8 & People obsessed with weight loss or anorexia.\\
 \hline
 6 & Expressing depression, suicidal tendencies, or social anxieties. \\
 \hline
 5 & Relate to Gay/Lesbian pornography. Most of the pictures are not tagged for nudity.\\
 \hline
 5 & Self-identifying as Female fitness or yoga accounts. Upload pictures of their diet or progress.\\
 \hline
 1 & Groups supporting Lesbian, Gay, Bisexual, Transgender and Queer (LGBTQ).\\
 \hline
80 & Miscellaneous\\
 \hline
\end{tabular}
\end{center}
%\vspace{-2.0em}
\end{table}

\subsection{Survey Evaluation}

As a final validation procedure, we asked 4 independent evaluators to examine the top 100 sensitive and the top 100 non-sensitive accounts (i.e. those farthest from the SVM hyperplane). The 200 accounts were presented to the evaluators in a random order. Each evaluator visited the 200 Twitter account pages and labeled each account as sensitive or non-sensitive based on their subjective judgement. The majority label was considered as the final label assignment for the account. 

Our automatic sensitive account classifier was largely consistent with the human labeling. For the 100 non-sensitive accounts, only 3 accounts were labeled by the human evaluators as sensitive. 
For the 100 sensitive accounts, 70 were labeled by the human evaluators as sensitive. For the remaining 30 accounts, 3 were protected accounts (activated the protected status after our initial data gathering), 13 self-identify as high school and college females, 7 self-identify as females posting about fitness, 2 self-identify as guys (a college student and an actor), and the remaining 5 were about sharing multimedia (pictures, Twitter Vine videos), love quotes, or not in English. From this survey we  see that the vast majority of accounts with a small fraction of anonymous followers and a large fraction of identifiable followers are indeed non-sensitive from the perspective of the subjective humans.  Also, the majority of the accounts with a large fraction of anonymous followers and a small fraction of identifiable followers are sensitive from the perspective of the subjective humans. However, there are accounts -- such as those maintained by high-school, college, and fitness-concerned females -- that many people would not consider to be sensitive but nevertheless have a large fraction of anonymous followers. 
This in itself is an interesting insight, and merits further investigation in the future. 

In summary, we have evaluated our methodology for discovering sensitive accounts from three different angles: LDA topic analysis using 2,000 sensitive and non-sensitive accounts; manually inspecting and categorizing the 300 most sensitive accounts; and finally, asking subjective humans to label 200 accounts as sensitive or non-sensitive. These evaluations clearly show that the vast majority of automatically-identified sensitive accounts are tweeting about topics that are very different from the automatically-identified non-sensitive accounts, that the majority of the  sensitive accounts deal with topics that most people consider to be sensitive
(including those that violate local laws), and that the methodology finds many sensitive topics -- such as obesity and anorexia - which were not originally hand-chosen in the prior study~\cite{PeddintiCOSN2014}. Finally the approach can easily scale across all of the Twitter accounts. All the experimentation was done on a single quad-core machine with 8GB RAM, and the sensitive account detection took just 1-2 hours to label the 93,000 accounts with 404 million followers.

\section{Related Work}
\label{sec:twittersensitiverelatedwork}

\subsection{Online Anonymity}

There is a huge body of research work based on surveys and user interviews that evaluates if online anonymity is harmful~\cite{LelkesJDCB2012,Millen2003} or beneficial~\cite{Joinson1999,PostmesSSd2001,ConnollyJV1990}.
A few have evaluated if users are actively seeking anonymity on the web~\cite{KangBK2013}, and if
anonymity features in the products are actively used~\cite{fourchan,PeddintiCOSN2014,Gomez2008}. 
In this paper, we do not focus on the debate of whether online anonymity is useful or harmful, or quantify if users exercise anonymity. 
We instead leverage user anonymity patterns to automatically detect sensitive accounts in social networks.

\subsection{Identifying Sensitive Content}

There has not been much work on exploring the diversity of topics that online users consider sensitive.  
Some attempts to understand user content sensitivity preferences have been restricted to capturing user views on a pre-determined list of topic categories~\cite{pewresearchsensitivity,hawkey2006examining,correa2015whisper}, or rely on user self-reporting during surveys and interviews~\cite{WangNKALC2011}. These methodologies have limitations of being subjective, or are expensive to capture across a sufficiently large user sample. 

Data-driven studies alleviate some of these limitations. 
\cite{correa2015whisper} analyzed 500 Whisper social media posts to understand the range of topics users consider sensitive.
A recently conducted data-driven study examines user content sensitivity preferences in the context of Quora, a question and answer service~\cite{Peddinti2014}. 
Though similarities exist to some of the inferences we draw in this paper and the Quora study, there are significant differences between the two. The Quora study deals with identifying sensitive topic categories, while we focus on identifying sensitive accounts. The Quora study relies on predefined content category tags,
while our study does not and is able to generalize to include overlooked topics.
Quora forces users to be completely identifiable~\cite{quorarealname}, and allows temporarily adopting complete anonymity (for example, when answering a sensitive question). The case of Twitter is more complex and noisy: (i) there are many different levels of user anonymity (Section~\ref{sec:userclassification}), and (ii) because the chosen anonymity is permanent, an anonymous user can follow both sensitive and non-sensitive accounts using the same profile.

\subsection{Other Studies on Twitter}
Researchers have analyzed the user social graph and performed tweet content analysis to understand: how users choose others to follow~\cite{Java2007}; identifying spam~\cite{twitterspamcharacterization,twitterspamsuspendedaccounts}; verifying  information credibility on social networks~\cite{Castillo2011}; 
determining information dissemination~\cite{twittersocialmedia}; 
measuring user influence~\cite{cha2010measuring}; identifying personally  identifiable information (PII) leakage~\cite{humphreys2010much} and developing machine learning classifiers to detect it~\cite{loosetweets}; detecting offensive tweets~\cite{offensivetweets}; and identifying individuals who are at risk for suicide~\cite{twittersuicidetracking}.
Unlike these existing studies, 
we rely on account follower anonymity preferences that, to the best of our knowledge, has not been previously explored.

\section{Conclusion}
\label{sec:conclusion}
We developed a novel methodology for identifying sensitive accounts on Twitter. Rather than using tweet word occurrences, our methodology is based on follower anonymity patterns. This approach not only easily scales across all Twitter accounts, but also is not limited to finding only accounts with pre-chosen words, and is not limited by the language features. To show that our methodology can process data at the Twitter scale, we applied it on a large Twitter crawl containing approximately 100,000 accounts with 404 million active followers.  

We evaluated our methodology three different ways. First we used LDA topic analysis using 2,000 sensitive and non-sensitive accounts.
The analysis clearly showed that  the identified sensitive accounts are tweeting about topics that are very different from the identified non-sensitive accounts. We then manually inspected and categorized the 300 most sensitive accounts, and observed that the majority of these accounts indeed tweet about topics that can be considered sensitive, including pornography, LGBTQ issues and depression. Finally, we asked subjective humans to label 200 accounts as sensitive or non-sensitive. The human labeling was largely consistent with our automated labeling.

Although our approach provides a scalable and objective way to understand content sensitivity, we believe more in-depth research is needed to improve user privacy preferences and expectations in the social media context. For instance, it is worth exploring and quantifying how many sensitive content categories are consistent across different social applications and how many are dependent on the nature of the application (e.g., photo sharing vs. messaging). Also, it needs to be seen if combining text based methods with user anonymity based methods can provide better identification rates. Our findings in this paper can help influence defining new privacy policies and controls. 

% Bibliography
\bibliographystyle{IEEEtran}
\bibliography{bibdata}

\end{document}